\documentclass[
reprint,
superscriptaddress,
showpacs,preprintnumbers,
 amsmath,amssymb,
 aps,
prx,
]{revtex4-1}

\usepackage{graphicx}
\usepackage{dcolumn}
\usepackage{bm}
\usepackage{amsmath}
\usepackage{esint}
\PassOptionsToPackage{normalem}{ulem}
\usepackage{ulem}
\usepackage{color}

\usepackage{geometry}
\begin{document}


\title{ Control of a quantum emitter's bandwidth by managing its reactive power }

\author{I\~nigo Liberal}
\author{I\~nigo Ederra}
\affiliation{Electrical and Electronic Engineering Department, Universidad P\'{u}blica de Navarra,
Campus Arrosad\'ia, Pamplona, 31006 Spain}
\affiliation{Institute of Smart Cities, Universidad P\'{u}blica de Navarra, Campus Arrosad\'ia,
Pamplona, 31006 Spain}
\author{Richard W. Ziolkowski}
\affiliation{Global Big Data Technologies Centre,University of Technology Sydney, Ultimo, NSW 2007,
Australia}

\date{\today}

\begin{abstract}
Reactive power plays a crucial role in the design of small antenna systems, but its impact on the
bandwidth of quantum emitters is typically disregarded. Here, we theoretically demonstrate that
there is an intermediate domain between the usual weak and strong coupling regimes where the
bandwidth of a quantum emitter is directly related to the dispersion properties of the reactive
power. This result emphasizes that reactive power must be understood as an additional degree of
freedom in engineering the bandwidth of quantum emitters. We illustrate the applicability of this
concept by revisiting typical configurations of quantum emitters coupled to resonant cavities and
waveguides. Analysis of the reactive power in these system unveils new functionalities, including
the design of efficient but narrowband photon sources, as well as quantum emitters exhibiting a
bandwidth narrower than its nonradiative linewidth.
\end{abstract}

\maketitle

\section{Introduction}

The analysis of the reactive power \cite{Harrington1961,Balanis1999} and related quantities such as
the stored energy \cite{Gustafsson2018,Yaghjian2005} plays a central role in the design of classical
radiating systems and the identification of their fundamental limits. This aspect is particularly
relevant for electrically small antennas, since the smaller the size of an emitter, the larger the
impact of the reactive fields on its performance. In fact, following the pioneering works of Wheeler
\cite{Wheeler1947}, Chu \cite{Chu1948} and Harrington \cite{Harrington1960}, much attention has been
devoted to the analysis of the stored energy and the derivation of physical bounds of antenna
performance
\cite{Collin1964,Fante1969,Mclean1996,Thal2006,Gustafsson2007,Yaghjian2010,Vandenbosch2011,Kim2016}
(see, e.g., \cite{Volakis2010} for a historical review). The importance of these works is that they
fundamentally establish what is possible and what is not possible to do with an antenna system. They
also inspire different antenna designs that approach the theoretical limits
\cite{Best2004,Ziolkowski2007,Ziolkowski2009,Kim2010}, and facilitate the implementation of
optimization procedures \cite{Gustafsson2013,Gustafsson2016,Jelinek2017}.

\begin{figure}[t!]
\centering
\includegraphics[width=3.35in]{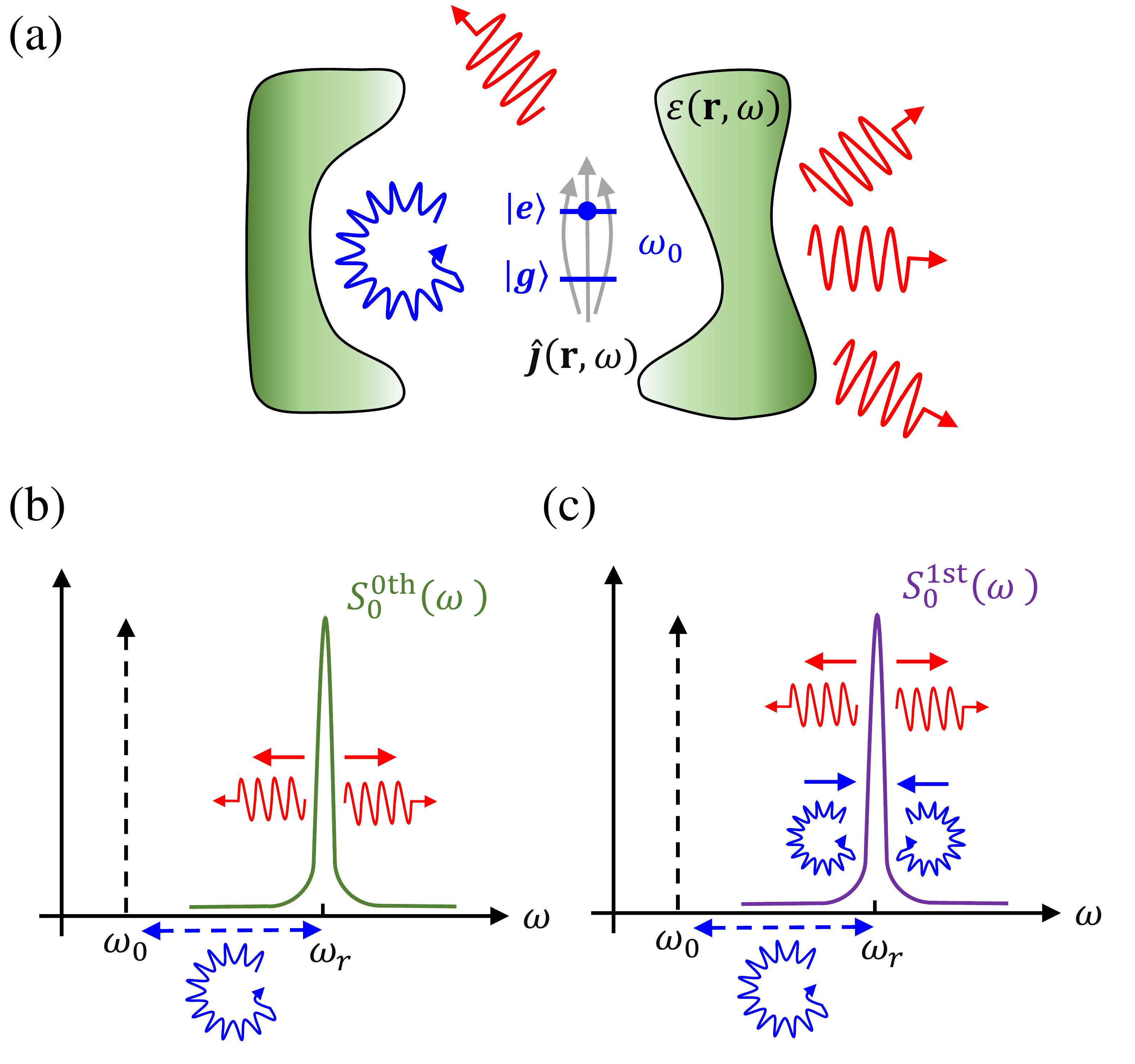}
\vspace{-0.1in}
\caption{(a) Sketch of the general configuration: A quantum emitter, modeled as a two-level system
$\left\{ \left|e\right\rangle ,\left|g\right\rangle \right\}$ with a transition frequency $\omega_0$,
has an effective current distribution
$\widehat{\mathbf{j}}\left(\mathbf{r},\omega\right)$ that
is coupled to a photonic environment characterized by a relative permittivity
$\varepsilon\left(\mathbf{r},\omega\right)$.
(b) Zeroth order (weak coupling) approximation to the emission spectrum, where reactive interactions
shift the emission frequency from $\omega_0$ to $\omega_r$, and radiative interactions define its
bandwidth.
(c) First-order correction to the spectrum where both radiative and reactive interactions impact the
emission bandwidth.}
\label{fig:Sketch}
\end{figure}

Quite to the contrary, the concept of reactive power is strange to the field of quantum optics and
the design of quantum emitters. Although interactions with so-called virtual photons are considered
(see, e.g., the recent perspective \cite{Scully2018}), these primarily lead to frequency shifts on
the emission frequency (see Figs.\,\ref{fig:Sketch}(a)-(b)). Different versions of these shifts
include the celebrated Lamb shift \cite{Lamb1947}, collective Lamb shift
\cite{Scully2009,Rohlsberger2010} and medium-assisted shifts
\cite{Hughes2009Lamb,Sokhoyan2013,Liberal2016}. In general, the spectrum is Lorentzian and its
linewidth is determined by the decay rate (see Fig.\,\ref{fig:Sketch}(b)). Therefore, it appears that
the interaction with virtual photons and/or reactive fields has no impact in the bandwidth of a
quantum emitter. This point might appear to be particularly surprising since most quantum emitters
are deeply subwavelength radiators, even more so than electrically small antennas.

At the same time, it is known that this behavior relates to the operation within the weak-copupling
regime, and, when a small quantum system is strongly coupled to a photonic nanostructure, their
interactions through the radiation field can significantly impact its emission spectrum. One
particularly popular example is the vacuum Rabi splitting, where the strong interaction between the
emitter and a cavity mode results in a two-peaked spectrum (see, e.g., \cite{Haroche2006}).
Therefore, it is clear that when the coupling is sufficiently strong, the energy stored in the
radiation field must have an impact of the bandwidth of a quantum emitter.

Here, we theoretically investigate the bandwidth of a quantum emitter amid its transition from the
weak to strong coupling regimes, emphasizing the role of the reactive power associated with the
emitter's current distribution. Specifically, we demonstrate that there is an intermediate domain
between the usual weak and strong coupling regimes in which the emission spectrum is Lorentzian, but
the associated bandwidth of the emitters is directly affected by the dispersion properties of the
reactive power (see Fig.\,\ref{fig:Sketch}(c)). This result highlights the reactive power as an
additional degree of freedom in controlling the quantum emitter's  bandwidth that can be harnessed to
introduce photon sources with unprecedented characteristics. Specifically, we will demonstrate how
managing the reactive power enables:
(i) increasing the efficiency of a quantum emitter while maintaining a narrow bandwidth, and,
(ii) designing a quantum emitter exhibiting a bandwidth narrower than its nonradiative linewidth.

\section{Theoretical framework}

As schematically depicted in Fig.\,\ref{fig:Sketch}, we investigate the emission properties of a
small quantum system modeled as a two-level system. It has excited
$\left|e\right\rangle =\int
d^{3}\mathbf{r}\,\psi_{e}\left(\mathbf{r}\right)\left|\mathbf{r}\right\rangle$
and ground
$\left|g\right\rangle =\int
d^{3}\mathbf{r}\,\psi_{g}\left(\mathbf{r}\right)\left|\mathbf{r}\right\rangle$ states that are
separated by the transition frequency $\omega_{0}$. The system is coupled to a macroscopic lossy
photonic environment that is characterized by the
dispersive relative permittivity
$\varepsilon\left(\mathbf{r},\omega\right)=\varepsilon_{R}\left(\mathbf{r},\omega\right)+i\,\varepsilon_{I}\left(\mathbf{r},\omega\right)$.
We model this quantum system within the framework of macroscopic QED (see, e.g., \cite{Vogel2006});
its Hamiltonian can be written as
\begin{equation}
\widehat{H}=\widehat{H}_{0}+\widehat{H}_{B}+\widehat{H}_{I}
\label{eq:H}
\end{equation}

\noindent with
\begin{equation}
\widehat{H}_{0}=\frac{\hslash\omega_{0}}{2}\,\widehat{\sigma}_{z}
\label{eq:H_0}
\end{equation}
\begin{equation}
\widehat{H}_{B}=
\int_{0}^{\infty}d\omega_{f}\int d^{3}\mathbf{r}
\,\hslash\omega_{f}\,\widehat{\mathbf{f}}^{\dagger}\left(\mathbf{r},\omega_{f}\right)\cdot\widehat{\mathbf{f}}\left(\mathbf{r},\omega_{f}\right)
\label{eq:H_B}
\end{equation}
\begin{equation}
\widehat{H}_{I}=\frac{1}{2m}\,\left[ \> \widehat{\mathbf{p}}-q\widehat{\mathbf{A}}\left(\widehat{\mathbf{r}}\right)\> \right]^{2}
\label{eq:H_I}
\end{equation}

\noindent where $\widehat{{\bf r}}$ and $\widehat{{\bf p}}$ are the position and momentum operators,
respectively, $m$ is the mass of the electron, $\widehat{\sigma}_{z}=\left|e\right\rangle
\left\langle e\right|-\left|g\right\rangle \left\langle g\right|$ and
$\widehat{\mathbf{f}}\left(\mathbf{r},\omega_{f}\right)$ are polaritonic operators representing the
excitations of the photonic environment. The vector potential operator is given by
\[
\widehat{\mathbf{A}}\left({\mathbf{r}}\right)=\int_{0}^{\infty}d\omega_{f}\,\int
d^{3}\mathbf{r}'\,\sqrt{\frac{\hslash}{\pi\varepsilon_{0}}}\,\frac{\omega_{f}}{c^{2}}\,
\sqrt{\varepsilon_{I}\left(\mathbf{r}',\omega_{f}\right)}
\]
\begin{equation}
\times
\left\lbrace
\mathbf{G}\left({\mathbf{r}},\mathbf{r}',\omega_{f}\right)\cdot\widehat{\mathbf{f}}\left(\mathbf{r}',\omega_{f}\right)
+ h.c.
\right\rbrace
\end{equation}

\noindent where $\mathbf{G}\left(\mathbf{r},\mathbf{r}',\omega_{f}\right)$ is the dyadic Green's
function of the macroscopic environment.

In order to draw a closer connection with classical antenna theory, we rewrite the interaction
Hamiltonian as a function of a current density operator. To this end, we disregard the
$\widehat{\mathbf{A}}^{2}\left({\mathbf{r}}\right)$ nonlinear term and expand the vector potential
operator in the position
representation to find that the interaction Hamiltonian can be rewritten as follows
\begin{equation}
\widehat{H}_{I}=-\int
d^{3}\mathbf{r}\,\,\widehat{\mathbf{j}}\left(\mathbf{r}\right)\cdot\widehat{\mathbf{A}}\left(\mathbf{r}\right)
\label{eq:Hint_jA}
\end{equation}

\noindent Here, we have defined the current density operator
\begin{equation}
\widehat{\mathbf{j}}\left(\mathbf{r}\right)=\frac{1}{2m}\,\,
\widehat{\rho}\left(\mathbf{r}\right)\,\widehat{\mathbf{p}}
+
h.c.
\label{eq:j_def}
\end{equation}

\noindent where
$\widehat{\rho}\left(\mathbf{r}\right)=q\left|\mathbf{r}\right\rangle \left\langle
\mathbf{r}\right|$. These operators are defined such that their expectation values recover the charge
density
$\rho\left(\mathbf{r},t\right)=\left\langle \widehat{\rho}\left(\mathbf{r}\right)\right\rangle
=q\left|\psi\left(\mathbf{r},t\right)\right|^{2}$
and the current density
$\mathbf{j}\left(\mathbf{r},t\right)=
\left\langle \widehat{\mathbf{j}}\left(\mathbf{r}\right)\right\rangle =
\frac{q}{2m}\,\left(-i\hslash\right)\psi^{*}\left(\mathbf{r},t\right)\nabla\psi\left(\mathbf{r},t\right)
+h.c.$,
in such a manner that they satisfy the continuity equation
$\partial_{t}\rho\left(\mathbf{r},t\right)+\nabla\cdot\mathbf{j}\left(\mathbf{r},t\right)=0$ (see,
e.g., in \cite{Binney2013} p. 32). For a two-level system, $\left\{ \left|e\right\rangle
,\left|g\right\rangle \right\} $,
the current density operator can be decomposed as follows:
$\widehat{\mathbf{j}}\left(\mathbf{r}\right)=
\mathbf{j}_{ge}\left(\mathbf{r}\right)\widehat{\sigma}+\mathbf{j}_{ge}^{*}\left(\mathbf{r}\right)\widehat{\sigma}^{\dagger}+\mathbf{j}_{ee}\left(\mathbf{r}\right)\widehat{\sigma}^{\dagger}\widehat{\sigma}+\mathbf{j}_{gg}\left(\mathbf{r}\right)\widehat{\sigma}\widehat{\sigma}^{\dagger}$,
with
$\mathbf{j}_{ab}\left(\mathbf{r}\right)=\left\langle
a\right|\widehat{\mathbf{j}}\left(\mathbf{r}\right)\left|b\right\rangle$ and
$\widehat{\sigma}=\left|g\right\rangle\left\langle e\right|$.

In the following, we will be mostly concerned with the properties of the fields generated by the
quantum emitters. Therefore, we compute the source field operators in the Heisenberg picture by
solving the equation of motion,
$i\hslash \, \partial_{t}\widehat{a}=\left[\widehat{a},\widehat{H}\right]$,
for the polaritonic operator $\widehat{\mathbf{f}}\left(\mathbf{r}',\omega_{f};t\right)$;
and we find that the Laplace transform of the source vector potential and electric field operator can
be conveniently written in analogy with their classical counterparts as functions of the current
density as follows
\begin{equation}
\widehat{\mathbf{A}}_{S}\left(\mathbf{r};\omega\right)=\mu_{0}\int
d^{3}\mathbf{r}'\,\mathbf{G}\left(\mathbf{r},\mathbf{r}',\omega\right)\cdot\widehat{\mathbf{j}}\left(\mathbf{r}',\omega\right)\label{eq:A_S}
\end{equation}
\begin{equation}
\widehat{\mathbf{E}}_{S}\left(\mathbf{r};\,\omega\right)=i\omega\mu_{0}\int
d^{3}\mathbf{r}'\,\mathbf{G}\left(\mathbf{r},\mathbf{r}',\omega\right)\cdot\widehat{\mathbf{j}}\left(\mathbf{r}',\omega\right)\label{eq:E_S}
\end{equation}

\section{Emission spectrum and reactive power}

Next we examine the emission spectrum during a decay process, i.e., when the emitter is initially
excited and the photonic environment is in its vacuum state: $\left|\psi\left(t=0\right)\right\rangle
=\left|e\right\rangle \left|\left\{ 0\right\} \right\rangle $. This configuration is relevant for
incoherent pumping or when the quantum emitter is resonantly excited via an initialization pulse.
Similar to the usual rotating wave approximation, we approximate the interaction Hamiltonian by
keeping only those terms that preserve the number of excitations:
\begin{equation}
\widehat{H}_{I}=
-\int d^{3}\mathbf{r}\,\left(
\widehat{\sigma}^{\dagger}\left(t\right)\,\mathbf{j}^{*}_{\mathrm{ge}}
\left(\mathbf{r}\right)\cdot\widehat{\mathbf{A}}_{S}^{\left(+\right)}\left(\mathbf{r};t\right)
+ h.c.\right)
\end{equation}

\noindent where $\widehat{\mathbf{A}}_{S}^{\left(+\right)}\left(\mathbf{r};t\right)$
is the inverse Laplace transform of
$\widehat{\mathbf{A}}_{S}^{\left(+\right)}\left(\mathbf{r};\omega\right)=\mu_{0}\int
d^{3}\mathbf{r}'\,\mathbf{G}\left(\mathbf{r},\mathbf{r}',\omega\right)\cdot\mathbf{j}_{\mathrm{ge}}\left(\mathbf{r}'\right)\widehat{\sigma}\left(\omega\right)$
and
$\widehat{\mathbf{A}}_{S}^{\left(-\right)}\left(\mathbf{r}; \omega \right)
=\left(\widehat{\mathbf{A}}_{S}^{\left(+\right)}\left(\mathbf{r};\omega\right)\right)^{\dagger}$.
Adapting the theory introduced in \cite{Wubs2004,Hughes2009,Hughes2012,Delga2014} to our current
density formulation within the one-photon correlation approximation, we find that the emission
spectrum is given by
\[
S\left(\mathbf{r},\omega\right)=\left\langle
\left(\widehat{\mathbf{E}}_{S}^{(+)}\left(\mathbf{r};\,\omega\right)\right)^{\dagger}\cdot\widehat{\mathbf{E}}_{S}^{(+)}\left(\mathbf{r};\,\omega\right)\right\rangle
\]
\begin{equation}
 =C_{\mathrm{prop}}\left(\mathbf{r},\omega\right)S_{0}\left(\omega\right)
 \label{eq:S_w}
\end{equation}

\noindent with
\begin{equation}
C_{\mathrm{prop}}\left(\mathbf{r},\omega\right)=\omega^{2}\mu_{0}^{2}\left|\int
d^{3}\mathbf{r}'\,\mathbf{G}\left(\mathbf{r},\mathbf{r}',\omega\right)\cdot\mathbf{j}_{\mathrm{ge}}\left(\mathbf{r}'\right)\right|^{2}
\label{eq:C_prop}
\end{equation}

\noindent being the propagation term which accounts for the directive emission properties of the
current density and its environment. The term $S_{0}\left(\omega\right)$ is the polarization
spectrum; it accounts for the impact of the emitter dynamics. It can be written as
\begin{equation}
S_{0}\left(\omega\right)=\left\langle
\widehat{\sigma}^{\dagger}\left(\omega\right)\widehat{\sigma}\left(\omega\right)\right\rangle
=\frac{1}{\left|\omega-\omega_{0}-\frac{1}{\hslash}\,\xi\left(\omega\right)\right|^{2}}
\label{eq:S0}
\end{equation}

\noindent We have defined in this expression the (in general dispersive) interaction energy term,
$\xi\left(\omega\right)$, which can be written as a function of the current densities as follows
\[
\xi\left(\omega\right)=\hslash\left[\> \triangle\omega\left(\omega\right)-i\,\frac{\Gamma\left(\omega\right)}{2}\>\right]
\]
\begin{equation}
=-\mu_{0}\int d^{3}\mathbf{r}\,\int d^{3}\mathbf{r}'\,
\mathbf{j}_{\rm
ge}^{*}\left(\mathbf{r}\right)\cdot\mathbf{G}\left(\mathbf{r},\mathbf{r}',\omega\right)\cdot\mathbf{j}_{\rm
ge}\left(\mathbf{r}'\right)
\label{eq:xi_w}
\end{equation}

We have identified the real and imaginary parts of  $\xi\left(\omega\right)$ with dispersive
frequency shift $\triangle\omega\left(\omega\right)$ and decay rate $\Gamma\left(\omega\right)$
terms, respectively. In addition, it is elucidating to draw an analogy with classical antenna theory.
To this end, we define $\mathbf{E}_{cl}\left(\mathbf{r},\omega\right)=i\omega\mu_{0}\int
d^{3}\mathbf{r}'\,\mathbf{G}\left(\mathbf{r},\mathbf{r}',\omega\right)\cdot\mathbf{j}_{\rm
ge}\left(\mathbf{r}'\right)$ as the classical time-harmonic field (${\rm exp}(-i\omega t)$
time-convention) that would be generated by the current distribution $\mathbf{j}_{\rm
ge}\left(\mathbf{r}'\right)$. In doing so, the interaction energy can be equivalently rewritten as
\cite{Balanis1999,Harrington1961}
\begin{equation}
\xi\left(\omega\right)=\frac{2}{i\omega}\,\left[\> P_{\mathrm{sup}}\left(\omega\right)+i\,P_{\mathrm{reac}}\left(\omega\right)\>\right]
\label{eq:xi_P}
\end{equation}

\noindent with
\begin{equation}
P_{\mathrm{sup}}=\frac{1}{2}\,\oint_{S_{\infty}}\,d\mathbf{S}\cdot\left(\mathbf{E}_{\mathrm{cl}}\times\mathbf{H}_{\mathrm{cl}}^{*}\right)
+\frac{\omega}{2}\,\int_{V_{\infty}}d^{3}\mathbf{r}\,\varepsilon_{0} \,
\varepsilon_{I}\left(\mathbf{r},\omega\right)\left|\mathbf{E}_{\mathrm{cl}}\right|^{2}
\label{eq:P_sup}
\end{equation}

\noindent and
\begin{equation}
P_{\mathrm{reac}}=\frac{\omega}{2}\,\int_{V_{\infty}}d^{3}\mathbf{r}\,\left[\> \varepsilon_{0} \,
\varepsilon_{R}\left(\mathbf{r},\omega\right)\left|\mathbf{E}_{\mathrm{cl}}\right|^{2}-\mu_{0}\left|\mathbf{H}_{\mathrm{cl}}\right|^{2}\>\right]
\label{eq:P_reac}
\end{equation}

\noindent The volume integrals are taken over an asymptotically large volume, $V_{\infty}$, bounded
by a surface $S_{\infty}$ in the far-zone of the sources $\mathbf{j}_{\rm
ge}\left(\mathbf{r}\right)$. On the one hand,
$P_{\mathrm{sup}}$ is the time-averaged power supplied by the current distribution $\mathbf{j}_{\rm
ge}\left(\mathbf{r}\right)$. It contains both the power radiated away from the system, as well as the
power dissipated in the surrounding environment. The reactive power, $P_{\mathrm{reac}}$, is related
to the energy stored in the electric and magnetic fields during the interaction process; but it does
not lead to any net energy transfer. However, it has a critical impact on the performance of
classical systems, including its bandwidth and robustness against undesired loss channels, as well as
stability and linearity aspects. Therefore, it can be concluded by comparing Eqs.~(\ref{eq:xi_w})
and (\ref{eq:xi_P}) that the dispersive decay rate $\Gamma$ is intimately related to the classical
supplied power $P_{\rm sup}$, while the frequency shift $\Delta\omega$ relates to the reactive power
$P_{\rm reac}$.

In general, it is clear from Eq.~(\ref{eq:S0}) that the polarization emission spectrum for a quantum
emitter is determined by the dispersion properties of the interaction energy term (\ref{eq:xi_w}). We
can expect that the spectrum will exhibit peaks at the resonant frequencies given by the solutions to
the implicit equation: $\omega_{r}=\omega_{0}+\triangle\omega\left(\omega_{r}\right)$. In the
neighborhood of one of these resonances, the zeroth order approximation to the emission spectrum
would be to neglect the dispersion properties of the spectrum around the resonant frequency, i.e.,
$\xi\left(\omega\right)\simeq\xi\left(\omega_{r}\right)$.
It recovers the usual Born-Markov approximation (or weak-coupling
regime), which leads to the zeroth order (Lorentzian) spectrum depicted in Fig. 1(b):
\begin{equation}
S_{0}^{\rm
0th}\left(\omega\right)=\frac{1}{\left(\omega-\omega_{r}\right)^{2}+\frac{\Gamma^{2}\left(\omega_{r}\right)}{4}}
\label{eq:S0_0th}
\end{equation}

\noindent Within this approximation, the 3dB bandwidth of the emission 
(frequency range between the half-maximum points)
will be simply given by the
decay rate $BW_{3dB}=\Gamma\left(\omega_{r}\right)$. In stark contrast with antenna theory, while the
reactive part of the interaction energy term is intimately related to the stored energy, it does not
have any impact on the quantum emitter's bandwidth. As anticipated at the outset, this feature is in
part surprising since quantum emitters are deeply subwavelength structures. Intuitively, the reason
for this behavior is that, in contrast with small antennas, the quantum emitter is already
intrinsically tuned to the resonance and the interaction with the electromagnetic field is considered
a small perturbation. Therefore, the reactive energy term only leads to a small perturbative
frequency shift.

However, this behavior changes when the strength of the coupling to the photonic environment is
increased and leads to significant changes on the emission spectrum. In order to elucidate the
transition between the usual weak and strong coupling regimes, we next introduce a first order
correction to the emission spectrum. To this end, we take a Taylor's series expansion of
$\xi\left(\omega\right)$ around $\omega_{r}$. We also note in analogy with resonant antennas
\cite{Yaghjian2005} that quantum emitters are typically tuned to be at either a maximum or a minimum
of the dispersive decay rate, i.e., at a frequency for which
$\partial_{\omega}\Gamma\left(\omega_{r}\right)\simeq0$.
The specific choice depends on whether one is using a photonic nanostructure to accelerate or
decelerate the spontaneous emission process. In this manner, the interaction energy term can be
approximated by
\begin{equation}
\frac{\xi\left(\omega\right)}{\hslash}\simeq\triangle\omega\left(\omega_{r}\right)+\left(\omega-\omega_{r}\right)\partial_{\omega}\triangle\omega\left(\omega_{r}\right)-i\,\frac{\Gamma\left(\omega_{r}\right)}{2}
\end{equation}

Under these conditions, a first-order correction to the emission spectrum can be written as
\begin{equation}
S_{0}^{\mathrm{1st}}\left(\omega\right)=A\,\frac{1}{\left(\omega-\omega_{r}\right)^{2}+\frac{1}{4}\left(\frac{\Gamma\left(\omega_{r}\right)}{1-\partial_{\omega}\triangle\omega\left(\omega_{r}\right)}\right)^{2}}
\label{eq:S0_1st}
\end{equation}

\noindent with $A=\left(1-\partial_{\omega}\triangle\omega\left(\omega_{r}\right)\right)^{-2}$. It is
clear from Eq.~(\ref{eq:S0_1st}) that the emission spectrum still preserves a Lorentzian lineshape
within this first-order correction. However, the linewidth is not entirely determined by the decay
rate, but it directly depends on the dispersion of the reactive part of the interaction energy term,
$BW_{3dB}=\Gamma\left(\omega_{r}\right)/(1-\partial_{\omega}\triangle\omega\left(\omega_{r}\right))$.
Therefore, we find in this intermediate domain that managing the reactive power and its dispersion
properties has a direct impact on the bandwidth of the emissions from nonclassical light sources (see Fig.\,\ref{fig:Sketch}(c)).

In view of this result, it is interesting to further draw analogies with classical antenna theory.
In particular, the dispersion properties of the frequency shift,
$\partial_{\omega}\triangle\omega\left(\omega\right)=\partial_{\omega}\left\{
2\omega^{-1}P_{\mathrm{reac}}\left(\omega\right)\right\} $, are directly related to the dispersion of
the reactive power, i.e., $\partial_{\omega}P_{\mathrm{reac}}\left(\omega\right)$. Adapting the
derivations in \cite{Yaghjian2005,Harrington1961} to our purposes, the latter can be written in terms
of field related quantities as follows
\[
\partial_{\omega}P_{\mathrm{reac}}=-\frac{1}{2}\,\int_{V_{\infty}}d^{3}\mathbf{r}\,\left[\> \mu_{0}\left|\mathbf{H}_{\mathrm{cl}}\right|^{2}
+\varepsilon_{0} \,
\partial_{\omega}\left\{\omega\varepsilon_{R}\left(\mathbf{r},\omega\right)\right\}
\left|\mathbf{E}_{\mathrm{cl}}\right|^{2} \>\right]
\]
\[
+2 \, \omega \, \varepsilon_{0}\int_{V_{\infty}}d^{3}\mathbf{r}\,\,\varepsilon_{I}\left(\mathbf{r},\omega\right)
\mathrm{Im}\left\{\mathbf{E}_{\mathrm{cl}}^{*}\cdot\partial_{\omega}\mathbf{E}_{\mathrm{cl}}\right\}
\]
\begin{equation}
+\mu_{0}\oint_{S_{\infty}}d\Omega\,r\left|\mathbf{F}\left(\mathbf{u}_{r},\omega\right)\right|^{2}
\label{eq:dw_Preac}
\end{equation}

\noindent where $\mathbf{F}\left(\mathbf{u}_{r},\omega\right)$ is the emission pattern in the
far-zone, i.e., in the limit
$\lim_{r\rightarrow\infty}\mathbf{E}_{\mathrm{cl}}\left(\mathbf{r},\omega\right)=\left(e^{i\frac{\omega}{c}r}/r\right)\,\mathbf{F}\left(\mathbf{u}_{r},\omega\right)$.

Equation (\ref{eq:dw_Preac}) provides information about the behavior of the system in some limiting
cases. For example, if the system can be considered lossless, i.e., when
$\varepsilon_{I}\left(\mathbf{r},\omega_{r}\right)\rightarrow0$, and nonradiating, i.e., when
$\mathbf{F}\left(\mathbf{u}_{r},\omega_{r}\right)\rightarrow0$ near the resonant frequency
$\omega_{r}$, we can write
$\partial_{\omega}P_{\mathrm{reac}}\left(\omega\right)\simeq
-\frac{1}{2}\,\int_{V_{\infty}}d^{3}\mathbf{r}\,\left[\> \mu_{0}\left|\mathbf{H}_{\mathrm{cl}}\right|^{2}+\varepsilon_{0}\partial_{\omega}\left\{
\omega\varepsilon_{R}\left(\mathbf{r},\omega\right)\right\}
\left|\mathbf{E}_{\mathrm{cl}}\right|^{2}\>\right]$. Consequently, the frequency derivative of the
reactive power will be negative, $\partial_{\omega}P_{\mathrm{reac}}\left(\omega\right)<0$, as a
manifestation of Foster's reactance theorem \cite{Harrington1961}. It can be readily checked that
$\partial_{\omega}\triangle\omega\left(\omega_{r}\right)<0$ for such a lossless and nonradiating
system. Actually, a similar behavior is expected in most cases since the reactive power is dominated
by contributions from the  near fields. This implies that taking into account the impact of the
reactive power will predict, in most cases, a narrower bandwidth of emission. However, for resonance
frequencies $\omega_{r}$ near strongly radiating and/or dissipative points it is possible to observe
the reverse behavior, i.e., $\partial_{\omega}P_{\mathrm{reac}}\left(\omega\right)>0$.
This outcome in turn leads to a broadening of the bandwidth.

In general, Eq.~(\ref{eq:S0_1st}) illustrates that there is an intermediate domain between the usual
weak and strong coupling regimes where the emission bandwidth can be controlled not only through the
decay rate, but also through the reactive power. This provides and additional degree of freedom in
controlling the bandwidth, which can be used to either broaden or compress it, and thus offers new
opportunities of engineering the emission spectrum of quantum emitters.

\section{Examples}

The basic theory introduced in the previous section can be applied to a variety of quantum emitters
and photonic nanostructures. In the following, we provide some examples illustrating the role of the
reactive power in typical configurations of quantum emitters coupled to photonic nanostructures. As
we will show, taking into account the role of the reactive power provides a better understanding of
the transition from the weak to the strong coupling regimes, and unveils novel functionalities even
in well-studied systems such as resonant cavities and waveguides.
\begin{figure*}[th]
\centering
\includegraphics[width=6in]{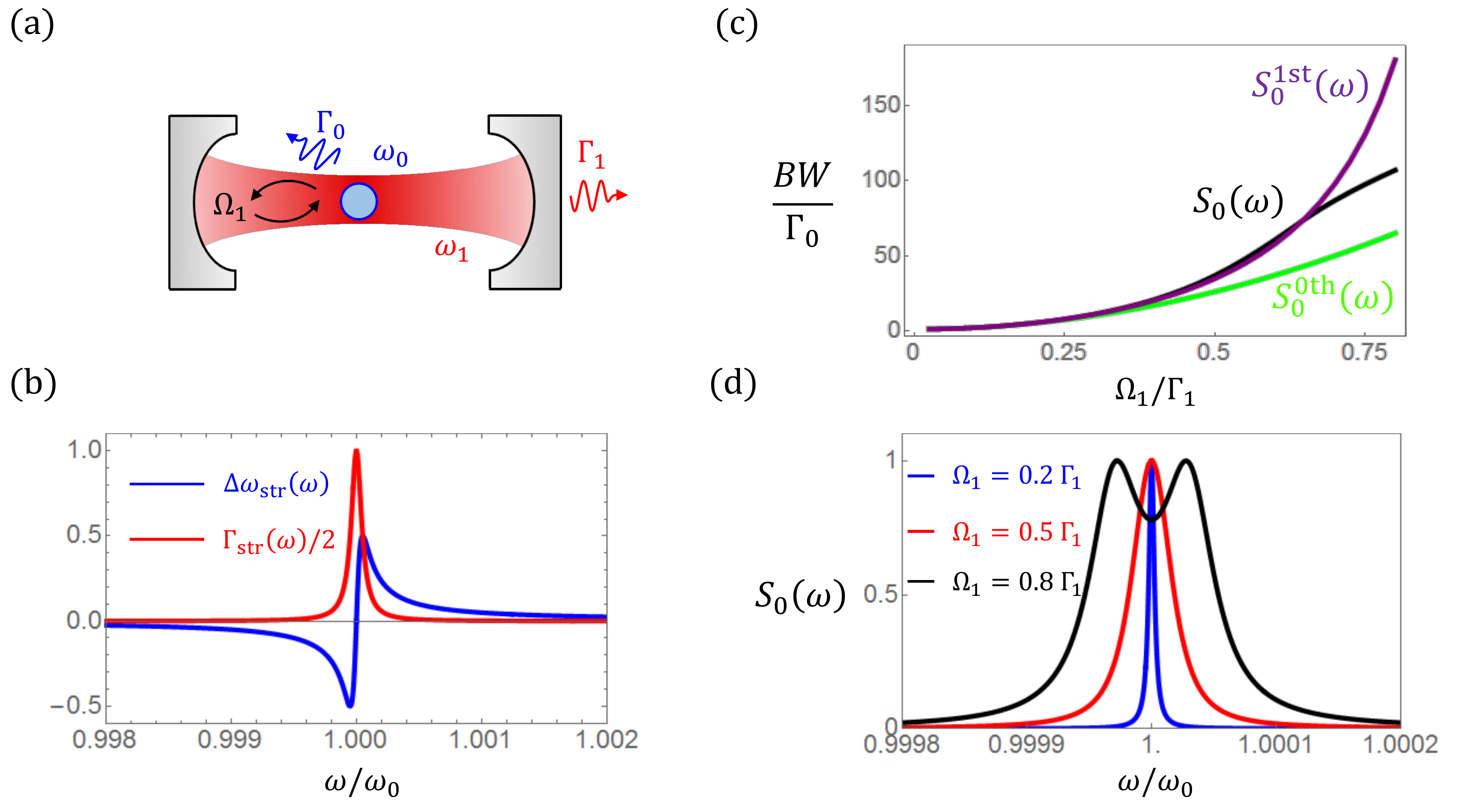}
\vspace{-0.1in}
\caption{(a) Sketch of the single-mode cavity geometry: A quantum emitter with transition frequency
$\omega_0$ and intrinsic decay rate $\Gamma_0=10^{-6}\omega_0$ is coupled to a single-mode cavity
with resonance frequency $\omega_1=\omega_0$, linewidth $\Gamma_1=10^{-4}\omega_0$, and coupling
strength $\Omega_1$. (b) Dispersive frequency shift $\Delta\omega_{\rm str}(\omega)$ and decay rate
$\Gamma_{\rm str}(\omega)$ normalized to their maximum value $\Omega_1^2/\Gamma_1$. (c) 3dB emission
bandwidth, normalized to the intrinsic decay rate $\Gamma_0$, as a function of the normalized coupling
strength $\Omega_1/\Gamma_1$. Comparison between the predictions for the full spectrum $S_0(\omega)$
and zeroth $S_0^{0{\rm th}}(\omega)$ and first $S_0^{1{\rm st}}(\omega)$ order approximations. (d)
Normalized emission spectrum for coupling strengths $\Omega_1=0.2\,\Gamma_1$,
$\Omega_1=0.5\,\Gamma_1$ and $\Omega_1=0.8\,\Gamma_1$.}
\label{fig:Fig2}
\end{figure*}

When considering the coupling of a quantum emitter to an inhomogeneous photonic environment, it is
convenient to decompose the dyadic Green's function
$\mathbf{G}\left(\mathbf{r},\mathbf{r}',\omega\right)=\mathbf{G}_{\rm
str}\left(\mathbf{r},\mathbf{r}',\omega\right)+\mathbf{G}_0\left(\mathbf{r},\mathbf{r}',\omega\right
)$ into the addition of a term associated to the modes of a structure of interest $\mathbf{G}_{\rm
str}\left(\mathbf{r},\mathbf{r}',\omega\right)$ (e.g., a cavity or a waveguide), as well as a term
$\mathbf{G}_0\left(\mathbf{r},\mathbf{r}',\omega\right)$ accounting for the rest of the optical
modes. Common decompositions include homogeneous and scattering parts \cite{Scheel1999,Dung2000} and
cavity and radiating modes \cite{Hughes2009}, although the decomposition into any arbitrary basis is
possible. This leads to a similar decomposition for the interaction energy term
$\xi\left(\omega\right)=\xi_{\mathrm{str}}\left(\omega\right)+\xi_{0}\left(\omega\right)$, where
$\xi_{\rm str}$ and $\xi_{0}$ correspond again to the contributions from the modes of the structure
of interest and the remaining modes not of interest, respectively. Subsequently, the polarization spectrum
can be written as follows:
\begin{equation}
S_0\left(\mathbf{r},\omega\right)=\frac{1}{\left|\omega-\omega_{0}+i\,\frac{\Gamma_{0}}{2}-\frac{1}{\hslash}\,\xi_{\mathrm{str}}\left(\omega\right)\right|^{2}}
\end{equation}

\noindent Here, we have assumed that the interactions with the modes not of interest represented by
$\xi_{0}\left(\omega\right)$ are in the weak coupling regime. Consequently, the transition frequency
$\omega_{0}$ is assumed to include a Lamb shift. Furthermore, we introduced an intrinsic decay rate
$\Gamma_{0}$ to account for all of the radiative decay paths different from the modes of interest.
This decay rate can also account for the nonradiative processes intrinsic to the emitter
\cite{Novotny2012,Hughes2009,Hughes2012,Delga2014}, although a more sophisticated description would
be required for nonradiative processes leading to an intrinsinc non-Lorentzian spectrum (e.g.,
emitters with large phonon sidebands).

\subsection{Single-mode cavity: transition from weak to strong coupling regimes}

We start by revisiting the popular example of coupling a quantum emitter with resonance frequency
$\omega_0$ to a high-Q single-mode cavity, characterized by the resonant frequency $\omega_{1}$,
linewidth $\Gamma_{1}$, and coupling strength distinguished by the vacuum Rabi frequency
$\varOmega_{1}$ (see Fig.\,\ref{fig:Fig2}(a)). For a high-Q cavity ($\omega_1\gg\Gamma_1$), the
interaction energy term associated to the cavity mode can be approximated by
\begin{equation}
\xi_{\mathrm{str}}\left(\omega\right)=\hslash\,\frac{\varOmega_{1}^{2}}{4}\frac{1}{\omega-\omega_{1}+i\frac{\Gamma_{1}}{2}}
\label{eq:xi_cav}
\end{equation}

\noindent This simple form of the interaction energy allows for a closed-form solution of the dynamic
response of the system. For example, the time evolution of the emitter population is characterized by
exponentially damped vacuum Rabi oscillations.

This formulation also provides a very clear picture of the impact of the reactive power on the
emission spectrum. In the following, we consider a narrowband emitter with $\Gamma_0=10^{-6}\omega_0$
that is coupled to a moderate Q resonator with $\Gamma_1=10^{-4}\omega_0$. This configuration is well
within the range of different optical cavities \cite{Vahala2003}. Fig.\,\ref{fig:Fig2}(b) represents
the decay rate $\Gamma_{\rm str}\left(\omega\right)$ and frequency shift $\Delta\omega_{\rm
str}\left(\omega\right)$, whose dispersion is characterized by the Lorentzian line of the cavity.
As anticipated, the frequency derivative of the frequency shift is negative at most frequencies,
i.e., $\partial_{\omega}\Delta\omega_{\rm str}\left(\omega\right)<0$. However, this trend is reversed
near the resonance:   $\omega\sim\omega_0$, where we observe $\partial_{\omega}\Delta\omega_{\rm
str}\left(\omega\right)>0$. Therefore, we can anticipate that as the coupling strength of an emitter
tuned to the cavity resonance is increased, the bandwidth will tend to be broadened with respect to
what could be expected from the zeroth order approximation (weak coupling regime) simply by looking
at the dispersion of the reactive term $\Delta\omega_{\rm str}\left(\omega\right)$. In this manner,
considering the impact of the dispersion of the reactive power provides additional insight into the
transition from the weak to the strong coupling regimes.

This point is more clearly illustrated in Fig.\,\ref{fig:Fig2}(c), which depicts the 3dB bandwidth as
a function of the coupling strength. It also compares the bandwidth predicted within the zeroth
(Eq.\,\ref{eq:S0_0th}) and first (Eq.\,\ref{eq:S0_1st}) order approximations. For small coupling
strengths: $\Omega_1 < 0.3 \,\Gamma_1$, the 3dB bandwidth is correctly predicted by all three
formulations. However, for larger coupling strengths, the common zeroth order approximation provides
a pessimistic prediction of the bandwidth, i.e., it fails to account for the broadening induced by
the dispersion of the reactive power. Our first-order correction correctly predicts the bandwidth for
an extended regime up to roughly $\Omega_1 \sim 0.7 \, \Gamma_1$. For larger coupling strengths, the
system enters into the strong coupling regime, and the spectrum is characterized by the well-known
two-peaked spectrum usually referred to as Rabi splitting
(see Fig.\,\ref{fig:Fig2}(d)).

\subsection{Two-mode cavity: highly-efficient narrowband source}

The information provided by the analysis of the reactive power can also be leveraged to introduce
photon sources with novel functionalities. We illustrate this point by analyzing a two-mode resonant
cavity and showing how this simple structure can be utilized to enable the design of
highly-efficient narrowband sources. Typically, the total decay rate of a quantum emitter is split
into the desired $\Gamma_{\rm str}(\omega_0)$ and undesired $\Gamma_0$ decay channels. The former
usually accounts for emission into a preferred optical mode, while the latter includes any
nonradiative decays and decays into unwanted photonic modes. Consequently, the emission efficiency
(quantum yield or beta factor) is defined as the ratio between the desired and total decay rates:
$\eta=\Gamma_{\rm str}(\omega_0)/(\Gamma_{\rm str}(\omega_0) + \Gamma_0)$ \cite{Novotny2012}.
Usually, efficient photon sources are designed by enhancing the decay rate of the desired channels by
means of coupling to photonic nanostructures, i.e., to ensure that $\Gamma_{\rm
str}(\omega_0)\gg\Gamma_0$. For this reason, increasing the efficiency is intrinsically associated
with bandwidth enlargement. In turn, this feature hinders the design of highly-efficient, but
narrowband photon sources.
\begin{figure*}[th]
\centering
\includegraphics[width=6in]{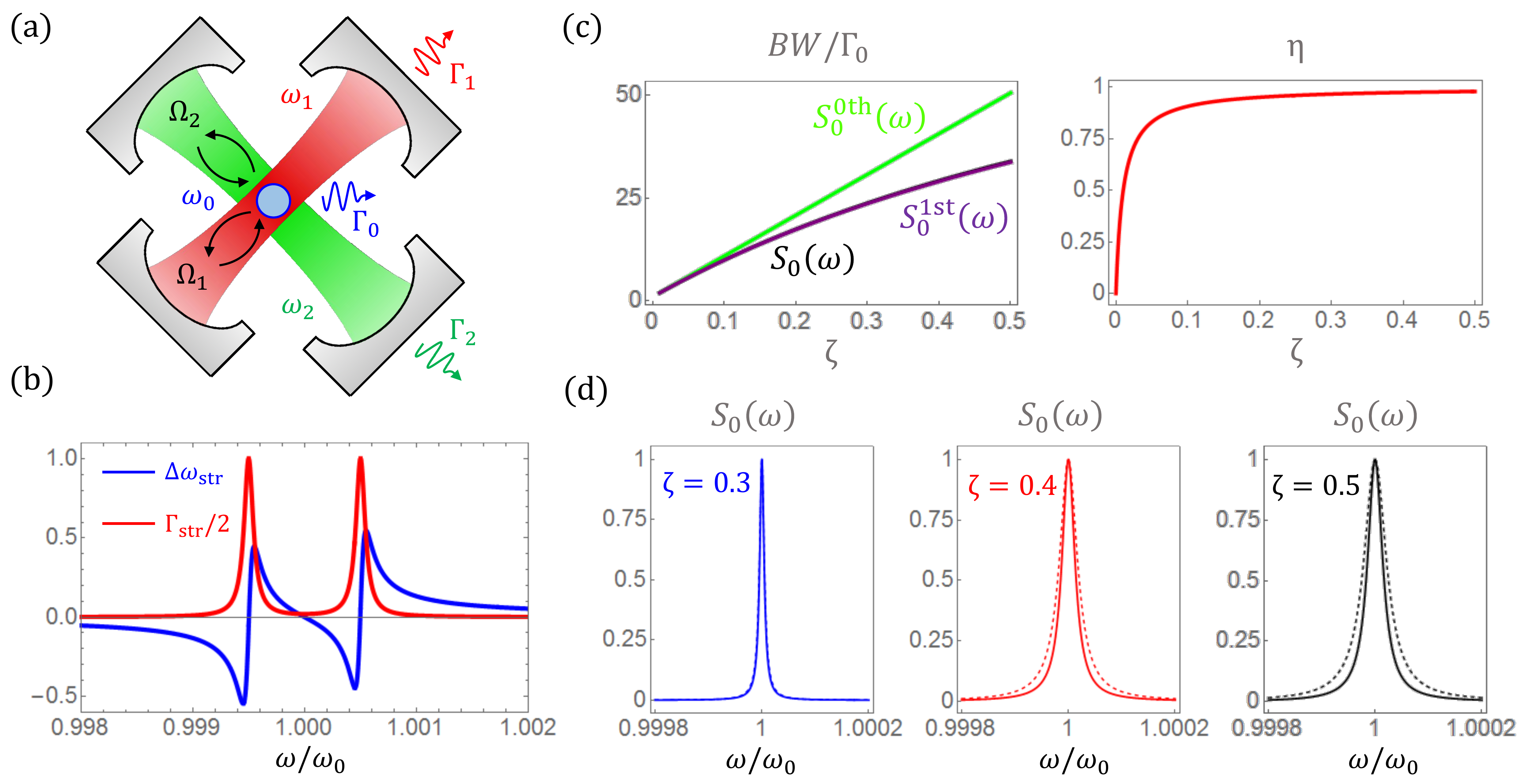}
\vspace{-0.1in}
\caption{(a) Sketch of the two-mode cavity geometry: A quantum emitter with transition frequency
$\omega_0$ and intrinsic decay rate $\Gamma_0=10^{-6}\omega_0$ is coupled to two single mode cavities
whose resonance frequencies are $\omega_1=\omega_0 + \omega_{\Delta}$ and $\omega_1=\omega_0 -
\omega_{\Delta}$. The detuning parameter $\omega_{\Delta}= 5\,\Gamma_1$, the linewidth
$\Gamma_1=\Gamma_2=10^{-4}\omega_0$, and the coupling strength $\Omega_1=\Omega_2$.
(b) Dispersive frequency shift $\Delta\omega_{\rm str}(\omega)$ and decay rate $\Gamma_{\rm
str}(\omega)$ normalized
to their maximum value $\Omega_1^2/\Gamma_1$. (c) Comparison between the predictions for the full
spectrum $S_0(\omega)$ and zeroth $S_0^{0{\rm th}}(\omega)$ and first $S_0^{1{\rm st}}(\omega)$ order
approximations. (Left) 3dB emission bandwidth normalized to the intrinsic decay rate $\Gamma_0$.
(Right) efficiency $\eta$ as a function of the coupling parameter
$\zeta=(\Omega_1/\omega_{\Delta})^2 \, / 2$. (d) Normalized spectrum for the coupling parameters $\zeta=0.3$,
$\zeta=0.4$ and $\zeta=0.5$. For reference, the zeroth order approximation is included as a dashed curve.}
\label{fig:Fig3}
\end{figure*}

However, highly-efficient, but narrowband single photon sources would be of great interest for a
number of applications. For example, narrowing the emission bandwidth would help in generating
indistinguishable photons, particularly when different physical systems are interfaced with their
sources \cite{Karpinski2017}. Narrowband sources would also facilitate spectroscopy with
non-classical light, either by interrogating biological or chemical samples with high spectral
precision, and/or by enhancing the emission from a molecular transition while avoiding the spectral
overlap with neighboring transitions. They would also expedite frequency-division multiplexing in
quantum communications.

Furthermore, managing the reactive power can provide a pathway to circumvent the compromise between
efficiency and narrowband operation. To illustrate this point, we consider a quantum emitter coupled
to a cavity supporting two non-interacting modes (or coupled to two different cavities)
as depicted in Fig. 3(a).
For the sake of simplicity, we assume that both resonant modes have similar characteristics in terms of coupling
strengths, $\Omega_1=\Omega_2$, and quality factors $\Gamma_1=\Gamma_2$, but their resonant
frequencies are detuned from the transition frequency of the emitter by symmetric shifts
$\omega_1=\omega_0-\omega_{\Delta}$ and $\omega_2=\omega_0+\omega_{\Delta}$, respectively. For this
configuration, the interaction energy term can be written as follows
\[
\xi_{\rm str}\left(\omega\right)=\hslash\,\frac{\varOmega_{1}^{2}}{4}
\left(
\frac{1}{\omega-\omega_{0}+\omega_{\Delta}+i\frac{\Gamma_{1}}{2}}
\right.
\]
\begin{equation}
\left.
+\frac{1}{\omega-\omega_{0}-\omega_{\Delta}+i\frac{\Gamma_{1}}{2}}
\right)
\label{eq:xi_cav2}
\end{equation}

The associated dispersion properties of the decay rate and frequency shift are depicted in
Fig.\,\ref{fig:Fig3}(b). These results show how the response of the system is characterized by the
superposition of two Lorentzian lines, each corresponding to one of the two uncoupled resonant modes.
Interestingly, we observe $\Gamma_{\rm str}\left(\omega_0\right)\simeq \zeta \,\Gamma_1$,
$\Delta\omega_{\rm str}\left(\omega_0\right)=0$, and
$\partial_{\omega}\Delta\omega_{\rm str}\left(\omega_0\right)\simeq -\zeta$ at the emitter transition
frequency, where we have defined the normalized coupling parameter
$\zeta = \left(\Omega_1/\omega_{\Delta}\right)^2$ \, / 2.
Therefore, this configuration allows for simultaneously enhancing the efficiency by increasing the
decay rate, while compressing the bandwidth by the action of the reactive power.

This effect is illustrated in Fig.\,\ref{fig:Fig3}(c), which depicts the 3dB bandwidth and efficiency
of the emitter as a function of the coupling factor $\zeta$. The figure shows that as the coupling
factor increases, the emission bandwidth becomes narrower than the one predicted within the zeroth
order approximation. For example, we have $\eta\simeq 0.98$ at $\zeta=0.5$, while exhibiting a
bandwidth
that is 33\% smaller than the one predicted purely based on the decay rate.
In this case, the first-order correction provides a very accurate prediction of the 3dB bandwidth for
the entire studied parameter range. This effect is justified by the fact that
$\partial_{\omega}^2\Delta\omega_{\rm str}\left(\omega_0\right)=0$, which substantially increases the
domain of validity of the first-order correction to the emission spectrum. The normalized spectrum for the coupling parameters $\zeta=0.3$,
$\zeta=0.4$ and $\zeta=0.5$ are reported in Fig.\,\ref{fig:Fig3}(d), which ratifies that the spectrum remains Lorentzian but with a bandwidth narrower than the prediction of the zeroth order approximation (shown as a dashed line).

It would be expected that additional functionalities will always come at some cost.
In this case, the efficiency achieved for a given cavity system will be smaller than if the emitter
was tuned at resonance with the cavity. However, once the quality of the cavity system is high enough
so that the efficiency at resonance would becomes saturated, our results demonstrate that one can
achieve a significant bandwidth compression while maintaining a high efficiency. In general, this
result sets the basis for the design of highly-efficient but narrowband single-photon sources.
Future evolutions of this concept might include many other configurations, for instance,
coupled-cavities and asymmetric systems, as well as the optimization of the involved parameters,
e.g., the quality factors and resonant frequencies of the cavities. These advanced design efforts are
beyond the scope of the present investigation.

\subsection{Multimode waveguide: sub-nonradiative linewidth}

The possibility of compressing the bandwidth by managing the reactive power poses the question of how
narrow the bandwidth of a quantum emitter could be theoretically. Typically, one can narrow the bandwidth
of a quantum emitter by reducing its decay rate (for example, by using a closed cavity
\cite{Kleppner1981} or a photonic crystal exhibiting a band-gap \cite{Bykov1972,Yablonovitch1987}).
However, both approaches come with the cost of sacrificing efficiency. Ultimately, this narrowing
process stops when the linewidth becomes dominated by nonradiative processes. However, the additional
degrees of freedom provided by the reactive power can allow us to circumvent this limit, potentially
getting access to sub-nonradiative loss linewidths.
\begin{figure*}[th]
\centering
\includegraphics[width=6in]{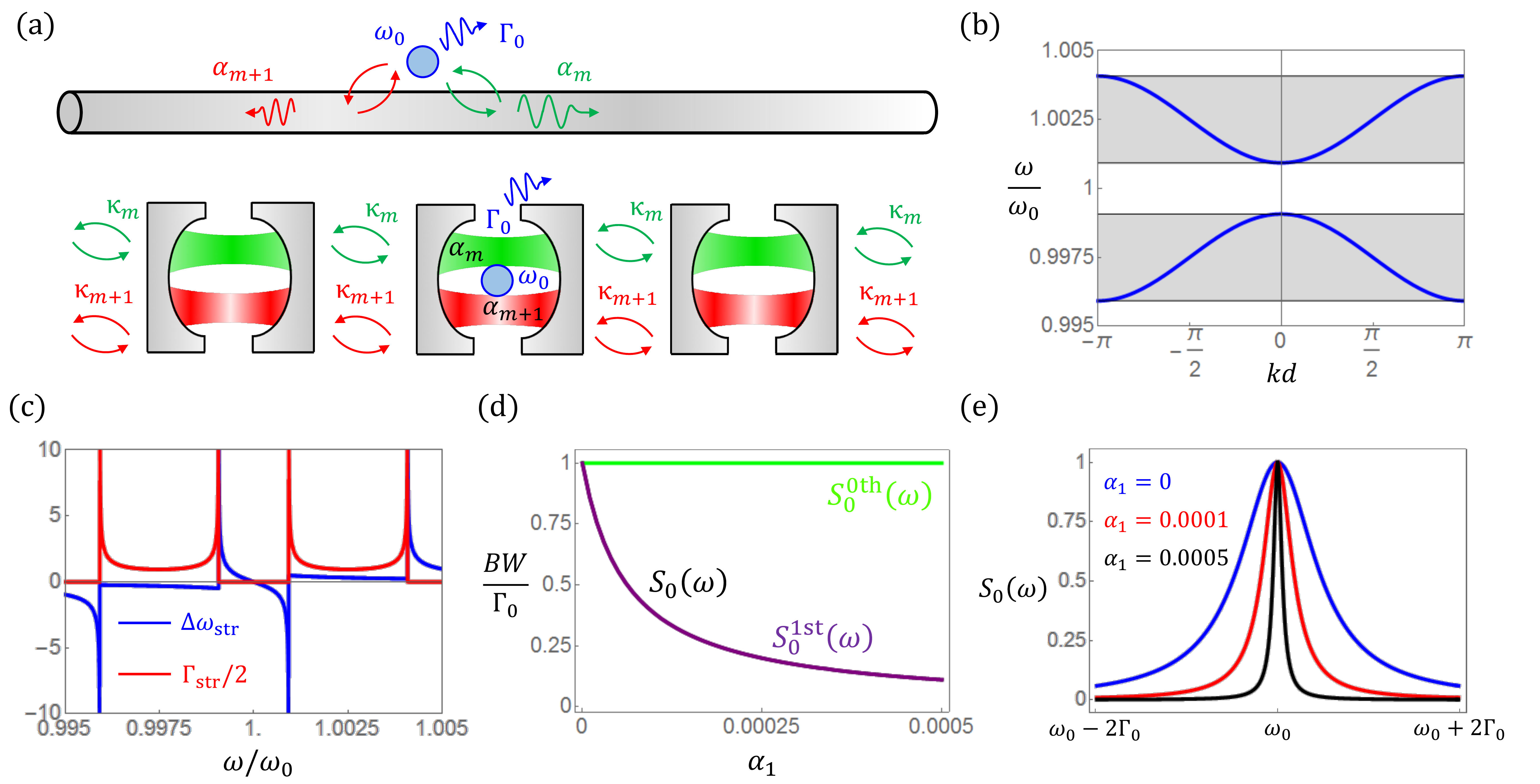}
\vspace{-0.1in}
\caption{(a) Sketch of the geometry: A quantum emitter, with transition frequency $\omega_0$ and
intrinsic decay rate $\Gamma_0$, decaying into $m$ modes of a photonic waveguide with
coupling factors $\alpha_m$. Potential implementation based on a coupled resonator optical waveguide
(CROW). (b) Dispersion diagram of the CROW. (c) Dispersive frequency shift $\Delta\omega_{\rm
str}(\omega)$ and decay rate $\Gamma_{\rm str}(\omega)$ normalized to its absolute value at the
center of the first band $\Gamma_{\rm str}(\omega_1)/2= \alpha_1 \, \omega_1 \, n_{g1}(\omega_1)$.
(d) 3dB emission bandwidth normalized to the intrinsic decay rate $\Gamma_0$ as a function of the coupling
parameter $\alpha_1=\alpha_2$. (e) Normalized spectrum for the coupling parameters $\alpha_1=0$, $\alpha_1=0.0001$
and $\alpha_1=0.0005$.}
\label{fig:Fig4}
\end{figure*}

We illustrate this possibility by examining a quantum emitter coupled to a multimode waveguide as
schematically depicted in Fig.\,\ref{fig:Fig4}(a). For this configuration, the interaction energy
term can be written as \cite{Lodahl2015,Hughes2004}
\begin{equation}
\xi_{\rm str}\left(\omega\right)= -i\,\hslash\omega \,\sum_{m}\,\, \alpha_m\,
n_{gm}\left(\omega\right)
\label{eq:xi_wg}
\end{equation}

\noindent where $\alpha_m$ is the coupling parameter to the $m^{th}$ mode. It includes, for instance,
the effects of the overlap of the emitter current distribution with the mode's field profile and its
effective volume. The group index of the $m^{th}$ mode is
$n_{gm}\left(\omega\right)= c/v_{gm}\left(\omega\right)$,
where $v_{gm}\left(\omega\right)$ is the associated group speed.

It is clear from Eq.\,(\ref{eq:xi_wg}) that engineering the dispersion properties of the group index
$n_{gm}\left(\omega\right)$ empowers the design of different light-matter interactions within optical
waveguides. To focus our discussion, we consider a coupled resonator optical waveguide (CROW)
illustrated in Fig.\,\ref{fig:Fig4}(a) \cite{Yariv1999,Poon2004}. A similar effect would be obtained
in other waveguides exhibiting a band-gap, such as photonic crystal \cite{Joannopoulos2011} and
metamaterial \cite{Caloz2005,Eleftheriades2005} waveguides.
The dispersion relation of a CROW waveguide within the tight-binding approximation can be described
as a set of $m$ pass-bands \cite{Yariv1999,Poon2004} whose individual dispersion relations
$\omega(k)=\omega_m + \kappa_m {\rm cos}(kd)$ are centered
around the resonance frequencies of the cavity $\omega_m$ and whose
bandwidths are equal to two times their coupling rates: $2\kappa_{m}$.
The group index associated with each of these modes is then given by
$n_{gm}(\omega) = n_{gm0}/\sqrt{1-(\omega-\omega_m)^2/\kappa_m^2}$,
with $n_{gm0}$ being the group index at the center of its pass-band \cite{Martinez2003}.

We consider the impact of two bands located around the emitters transition frequency. We set
$\omega_1=0.9975\,\omega_0$, $\omega_2=1.0025\,\omega_0$, $\kappa_1=\kappa_2=0.00158\,\omega_0$ and
$n_{gm0}=15$. As depicted in Fig.\,\ref{fig:Fig4}(b), a band structure and group index similar to
those reported in \cite{Cooper2010} are obtained.

Fig.\,\ref{fig:Fig4}(c) presents the corresponding dispersive frequency shift
$\Delta\omega_{\rm str}(\omega)$ and decay rate $\Gamma_{\rm str}(\omega)$.
They serve to illustrate some of the salient features of
the light-matter interactions within dispersive waveguides. For example, the decay rate is strongly
enhanced near the edges of the pass-bands since it is associated with a large group index, i.e., a
near-zero group velocity  \cite{Lodahl2015}. Similarly, the medium-assisted Lamb shift is enhanced at
the side of the band-edge that lies within the band-gap \cite{Sokhoyan2013}. On the other hand, the
decay rate is strongly suppressed within the band-gaps, leading to an inhibition of the spontaneous
emission \cite{Bykov1972,Yablonovitch1987} and the formation of long-lived bound states
\cite{John1990,Calajo2016,Shi2016}.

Simultaneously, Fig.\,\ref{fig:Fig4}(c) suggests new opportunities associated with the management of
the reactive power within the band-gap. For example, we note that if an emitter tuned within the
band-gap has a non-zero intrinsic decay rate $\Gamma_0$, then the dynamics of the quantum emitter
would still be dominated by an exponential relaxation through the channels external to the waveguide
system. This feature is true even if $\Gamma_{\rm str}(\omega_0)=0$. In such a case, the emission
spectrum would be expected to be a Lorentzian line with a 3dB bandwidth $\Gamma_0$. However, at the
center of the band-gap, we have $\Delta\omega_{\rm str}(\omega)=0$ and a negative frequency
derivative $\partial_{\omega}\Delta\omega_{\rm str}(\omega)<0$. These are the necessary ingredients
for bandwidth compression beyond that induced by an inhibition of spontaneous emission. This effect
is shown in Figs.\,\ref{fig:Fig4}(d) and \ref{fig:Fig4}(e) in which the quantum emitter 3dB bandwidth
and emission spectrum are depicted as functions of the coupling parameter $\alpha_1=\alpha_2$.
As expected, the bandwidth is identical to the intrinsic decay rate $\Gamma_0$ for small
coupling parameters $\alpha_1\sim 0$. On the other hand, it is rapidly compressed beyond this limit as
the coupling parameter strengthens and the zeroth order approximation is no longer valid. Again,
different parameters of the system, e.g., the separation and width of the propagating bands, could be
optimized to achieve a better performance for specific waveguide implementations and/or particular
applications.

In general, these results demonstrate the real possibility of using a photonic nanostructure to purely
compress the bandwidth of a quantum emitter beyond the limit of its nonradiative linewidth.
It is worth remarking that the so-called sub-natural linewidth photon emission based on
resonance fluorescence operating in the Heitler's regime has been reported
\cite{Heitler1954,Nguyen2011,Matthiesen2012}.
However, recent theoretical developments indicate that subnatural-linewidth and antibunching cannot
be observed simultaneously in this configuration unless the coherent part of the emitted light is
reduced by destructive interference with an external coherent signal \cite{Carreno2018}. The
operating principle of our configuration is entirely different. First, since it is not based on
resonance fluorescence, it does not require the exact compensation of different terms in order to
guarantee antibunching. Second, our proposed system is consistent with incoherent pumping and is thus
compatible with electronically-driven devices. In fact, if the intrinsic decay rate $\Gamma_0$ is
dominated by a radiative component (outside the waveguide system), our system would allow for
on-demand operation. Finally, achieving an intrinsic line narrower than the width associated with
the nonradiative losses might have important implications in the
dynamics of different decoherence channels beyond manipulating the emission bandwidths of quantum emitters.

\section{Conclusions}

Our results demonstrate that reactive power must be considered as an additional degree of freedom in
controlling the bandwidth of quantum emitters. As demonstrated, the impact of the reactive power can
result either in the compression or expansion of the bandwidth of a quantum emitter, simply depending
on the sign of its frequency derivative. Being able to control the bandwidth of emission beyond the
manipulation of its decay rate provides a finer control and offers new possibilities.
On the one hand, it is possible to elude a direct relationship between the bandwidth and efficiency.
This feature facilitates the design of efficient quantum emitters preserving a narrow bandwidth.
It also enables the compression of the source's bandwidth beyond limits imposed by the decay rates
intrinsic to the emitter.
We have outlined the basic theory and presented examples associated with applications involving
resonant cavities and waveguides. This basic theory and these configurations could be implemented
through a variety of systems, including different quantum emitters (cold atoms, ions, quantum dots,
color centers, ... ), multiple emitters, resonant cavities (defect cavity modes in photonic crystals,
nanopillar cavities, whispery gallery modes, plasmonic cavities, ...) and/or photonic crystal and
metamaterial waveguides. Many other configurations. e.g., coupled cavities, nanoparticle systems, and
waveguides with different dispersion profiles, could also be explored. In general, our results take
inspiration from classic antenna theory to provide a new perspective on the interactions of
quantum emitters with their photonic environments. Moreover, they may find important applications
in the development of nonclassical light sources.

\vspace{0.1in}

\begin{acknowledgments}
I.L. acknowledges support from a Juan de la Cierva – Incorporaci\'on Fellowship and dotaci\'on
adicional Caixa.
\end{acknowledgments}

\vspace{-0.2in}

\bibliography{library}

\vfill
\end{document}